\documentclass[sigconf]{acmart}
\AtBeginDocument{%
  }

\setcopyright{acmlicensed}
\copyrightyear{2026}
\acmYear{2026}
\acmDOI{XXXXXXX.XXXXXXX}
\acmConference[WWW '26]{the ACM on Web Conference 2026}{April 13--17,
  2026}{Dubai, UAE}
\acmISBN{978-1-4503-XXXX-X/2018/06}

\usepackage{algorithm}
\usepackage{algorithmic}

\usepackage{newfloat}
\usepackage{listings}

\usepackage{graphicx}
\usepackage{xspace}
\usepackage{url}
\usepackage{booktabs}

\usepackage{pifont}
\usepackage{amssymb}
\usepackage{multirow}
\usepackage{amsthm}
\usepackage{makecell}
\newtheorem{definition}{Definition}
\usepackage{amsmath}
\usepackage{tikz}
\usepackage{tcolorbox}
\usepackage{xcolor}
\usepackage{hyperref}
\newcommand{\system}{{TAGFN}\xspace}
\newcommand{\cmark}{\text{\ding{51}}}%
\newcommand{\xmark}{\text{\ding{55}}}%

\begin{document}

\title{\system: A Text-Attributed Graph Dataset for Fake News Detection in the Age of LLMs}

\author{Kay Liu}
\orcid{0000-0002-2022-9465}
\affiliation{
  \institution{University of Illinois Chicago}
  \city{Chicago}
  \state{Illinois}
  \country{USA}
}
\email{zliu234@uic.edu}

\author{Yuwei Han}
\orcid{0000-0001-8399-5469}
\affiliation{
  \institution{University of Illinois Chicago}
  \city{Chicago}
  \state{Illinois}
  \country{USA}
}
\email{yhan86@uic.edu}

\author{Haoyan Xu}
\orcid{0009-0008-7691-2926}
\affiliation{
  \institution{University of Southern California}
  \city{Los Angeles}
  \state{California}
  \country{USA}
}
\email{haoyanxu@usc.edu}

\author{Henry Peng Zou}
\orcid{0009-0003-5259-4998}
\affiliation{
  \institution{University of Illinois Chicago}
  \city{Chicago}
  \state{Illinois}
  \country{USA}
}
\email{pzou3@uic.edu}

\author{Yue Zhao}
\orcid{0000-0003-3401-4921}
\affiliation{
  \institution{University of Southern California}
  \city{Los Angeles}
  \state{California}
  \country{USA}
}
\email{yue.z@usc.edu}

\author{Philip S. Yu}
\orcid{0000-0002-3491-5968}
\affiliation{
  \institution{University of Illinois Chicago}
  \city{Chicago}
  \state{Illinois}
  \country{USA}
}
\email{psyu@uic.edu}

\renewcommand{\shortauthors}{Liu et al.}

\begin{abstract}
Large Language Models (LLMs) have recently revolutionized machine learning on text-attributed graphs, but the application of LLMs to graph outlier detection, particularly in the context of fake news detection, remains significantly underexplored. One of the key challenges is the scarcity of large-scale, realistic, and well-annotated datasets that can serve as reliable benchmarks for outlier detection. To bridge this gap, we introduce \system, \textit{a large-scale, real-world text-attributed graph dataset for outlier detection}, specifically fake news detection. \system enables rigorous evaluation of both traditional and LLM-based graph outlier detection methods. Furthermore, it facilitates the development of misinformation detection capabilities in LLMs through fine-tuning. We anticipate that \system will be a valuable resource for the community, fostering progress in robust graph-based outlier detection and trustworthy AI. The dataset is publicly available at \url{https://huggingface.co/datasets/kayzliu/TAGFN} and our code is available at \url{https://github.com/kayzliu/tagfn}.
\end{abstract}




\received{20 February 2007}
\received[revised]{12 March 2009}
\received[accepted]{5 June 2009}

\maketitle

\section{Introduction}
Graph-structured data offers a flexible framework for modeling interactions among entities in diverse domains such as social media networks \cite{xiao2020timme, liu2023data}. In many practical applications, nodes are often associated with rich textual attributes, forming text-attributed graphs (TAGs) \cite{yang2021graphformers, yan2023comprehensive}. By integrating structural and semantic information, TAGs enable more fine-grained learning.

\begin{figure}[t]
    \centering
    \includegraphics[width=0.8\linewidth]{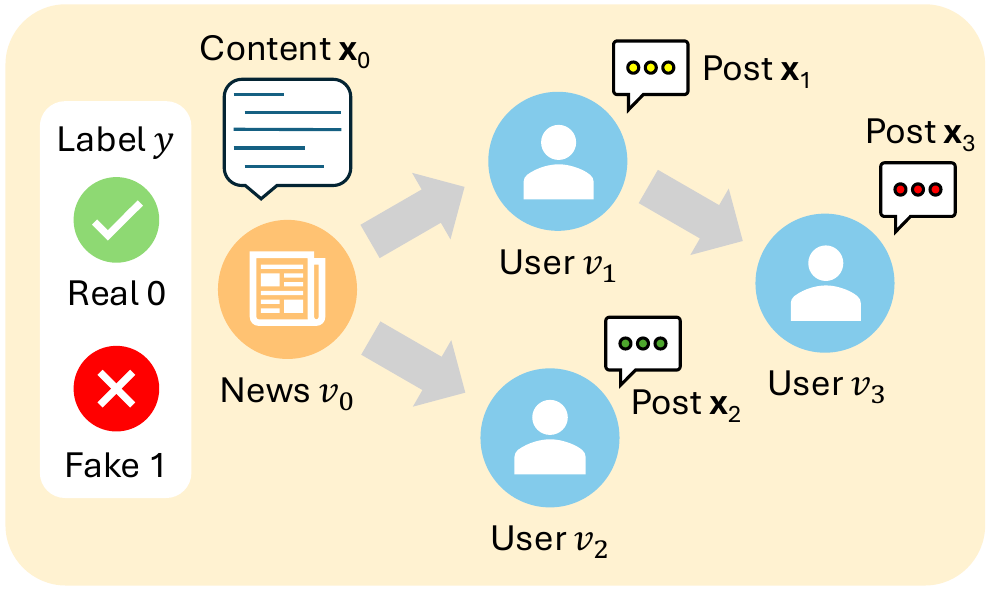}
    \caption{A toy example of news propagation graph in \system, where the root node denotes the news and child nodes represent users, each attributed with text.}
    \label{fig:example}
    \vspace{-0.15in}
\end{figure}

A growing body of work has explored the integration of graph learning with large language models (LLMs) for TAGs, yielding impressive results on tasks such as node classification \cite{chen2023label, zhu2025graphclip}, out‑of‑ distribution detection \cite{xu2025graph}, and question answering \cite{yasunaga2022linkbert, li2025taco}. Among the diverse applications of outlier detection, misinformation detection emerges as a natural use case for TAGs. The semantical information in news and user content, combined with the propagation graph structure, collectively provides critical signals for identifying fake news. Despite the promise of LLMs for graph learning \cite{li2023survey, chen2024exploring}, their application to outlier detection on graphs \cite{liu2022bond}--particularly for fake news detection and misinformation detection--remains largely underexplored. Most existing outlier detection methods are developed for graphs without textual information, leaving outlier detection on TAGs significantly understudied. 

In this context, the synergy between the graph structure and textual attributes plays a critical role in understanding phenomena such as information dissemination and the emergence of outliers (e.g., fake news). However, a primary obstacle to investigating outlier detection on TAGs is \textit{the scarcity of large-scale, realistic, and well-annotated datasets that combine graph structure with meaningful textual attributes and reliable ground‑truth labels}. 
Existing benchmarks are limited in scale, lack raw textual attributes, or fail to provide realistic ground-truth labels necessary for the rigorous evaluation of outlier detection methods.

To bridge this gap, we introduce \system, a large-scale, real-world TAG dataset for outlier detection in the context of fake news detection. As shown in Figure \ref{fig:example}, each graph in \system depicts the propogation of a news, where the root node represents the news itself and child nodes represent the users who propagated it. Each node is attributed with text--either the news content or user posts--capturing the multifaceted nature of information disseminations. We also provide ground-truth outlier labels for each graph/news, indicating whether the news is fake or real. We anticipate that \system will serve as a valuable resource for the research community, catalyzing progress at the intersection of LLMs, graph learning, and misinformation detection. By enabling systematic evaluation and development of advanced models, \system aims to advance the state of the art in both graph machine learning and trustworthy AI. 

Our contributions are mainly as follows:
\begin{itemize}
\item We construct \system, the first large-scale, real-world TAG dataset for outlier detection in the domain of fake news detection, addressing a critical gap in existing resources.
\item We provide baseline experiments to facilitate rigorous evaluation and comparison of graph learning and LLM-based fake news detection approaches.
\item We release the dataset and code, fostering further research in robust graph outlier detection and trustworthy AI.
\end{itemize}

\section{Related Work}

In this section, we review related datasets to our work. We compare \system with existing datasets in Table~\ref{tab:comparison} along the following aspects: presence of raw \textbf{Text} attributes, task of \textbf{Outlier} detection, scale 1M+ nodes/rows (\textbf{Large}), inclusion of \textbf{Graph} structure, and availability of \textbf{Time} information.

\begin{table}[t]
\centering
\caption{Comparison of \system with existing datasets.}
\label{tab:comparison}
\scalebox{0.7}{
\begin{tabular}{@{}l|ccccc@{}}
\toprule
            & \makecell{UPFD \\ \citeyearpar{dou2021user}} & \makecell{CS-TAG \\ \citeyearpar{yan2023comprehensive}} & \makecell{AD-LLM \\ \citeyearpar{yang2024ad} } & \makecell{NLP-ADBench \\ \citeyearpar{li2024nlp}} & \system \\ \midrule
Text        & \textcolor{purple}{\xmark} & \textcolor{teal}{\cmark}    & \textcolor{teal}{\cmark}     & \textcolor{teal}{\cmark}     & \textcolor{teal}{\cmark}     \\
Outlier     & \textcolor{teal}{\cmark}   & \textcolor{purple}{\xmark}  & \textcolor{teal}{\cmark}     & \textcolor{teal}{\cmark}     & \textcolor{teal}{\cmark}     \\
Large       & \textcolor{purple}{\xmark} & \textcolor{teal}{\cmark}    & \textcolor{purple}{\xmark}   & \textcolor{purple}{\xmark}   & \textcolor{teal}{\cmark}     \\
Graph       & \textcolor{teal}{\cmark}   & \textcolor{teal}{\cmark}    & \textcolor{purple}{\xmark}   & \textcolor{purple}{\xmark}   & \textcolor{teal}{\cmark}     \\
Time        & \textcolor{purple}{\xmark} & \textcolor{purple}{\xmark}  & \textcolor{purple}{\xmark}   & \textcolor{purple}{\xmark}   & \textcolor{teal}{\cmark}     \\ \bottomrule
\end{tabular}}
\end{table}

\subsection{Text-Attributed Graph Datasets}

Text-attributed graph (TAG) datasets have become a cornerstone for advancing research at the intersection of graph learning and LLMs. The recent surge in interest has led to the development of a diverse array of benchmarks. \citet{yan2023comprehensive} introduce CS‑TAG, a diverse and large-scale suite of benchmark datasets for TAGs, and establish standardized evaluation protocols. Recognizing the importance of temporal dynamics, \citet{NEURIPS2024_a65d054a} introduced DTGB, a large-scale dynamic TAG dataset. \citet{10.5555/3737916.3739865} further proposed TEG-DB, which incorporates both node and edge textual attributes. Despite these advances, there is no real-world TAG dataset for outlier detection.

\subsection{Fake News Detection Datasets}

Despite the proliferation of fake news detection datasets, most existing resources primarily focus on news content and basic metadata. \citet{wang2017liar} release is early fake new detection dataset LIAR, which includes PolitiFact‑annotated short statements with rich meta‑data (truthfulness, speaker, context, party affiliation, etc.) \citet{nakamura2019r} introduces Fakeddit, a multimodal fake news dataset of Reddit posts with paired text, images, and some metadata. FakeNewsNet \citep{shu2020fakenewsnet} extends the LIAR by assembling multi-dimensional social media‐based data, integrating full news content, social context, and spatiotemporal diffusion patterns. UPFD \citep{dou2021user} further integrate user profiles and propagation graphs, enabling the study of social dynamics in fake news dissemination. However, UPFD lacks raw textual attributes, limiting its flexibility for LLM-based fake news detection. 
Despite recent advances in LLMs and outlier detection—such as AD-LLM \citep{yang2024ad} and NLP-ADBench \citep{li2024nlp}—existing approaches remain primarily focused on textual content.
To bridge the gap, \system offer not only news content, but also the propagation graph structure and user historical posts content, providing a more holistic view the detection of fake news.

\section{\system}

\system is a text-attributed graph (TAG) dataset for graph level outlier detection in the domain of fake news detection. We present three subsets of varying scales: Politifact, Gossicop, and Fakeddit. Table~\ref{tab:stats} summarizes their statistics, including the number of nodes, edges, and graphs; the average graph size (in number of nodes); fake news ratio; and split size (train/validation/test) in graph count.

\begin{table}[t]
  \centering
  \caption{Statistics of the three subsets of \system.}
  \label{tab:stats}
  \scalebox{0.9}{
  \begin{tabular}{@{}c|ccc@{}}
    \toprule
    & \textbf{Politifact} & \textbf{Gossipcop} & \textbf{Fakeddit} \\
    \midrule
    \# Nodes    & 41,054     & 314,262    & 7,249,803 \\
    \# Edges    & 40,740     & 308,798    & 6,683,699 \\
    \# Graphs   & 314        & 5,464      & 566,104   \\
    Avg. Size   & 131        & 58         & 13        \\
    Fake (\%)   & 50.0       & 50.0       & 59.6      \\
    Train       & 62         & 1,092      & 467,538   \\
    Validation  & 31         & 546        & 49,186    \\
    Test        & 221        & 3,826      & 49,380    \\
    \bottomrule
  \end{tabular}}
\end{table}

\begin{table*}[t]
  \begin{minipage}[t]{0.49\textwidth}
    \captionof{table}{Performance across supervision levels (\%).}
    \centering
    \label{tab:main}
    \scalebox{0.74}{
    \renewcommand{\arraystretch}{1.15}
    \setlength{\tabcolsep}{7pt}
    \begin{tabular}{l|cc|cc|cc}
      \toprule
      \multirow{2}{*}{\textbf{Method}} 
        & \multicolumn{2}{c|}{\textbf{Politifact}} 
        & \multicolumn{2}{c|}{\textbf{Gossipcop}} 
        & \multicolumn{2}{c}{\textbf{Fakeddit}} \\
      & {ACC} & F1 
      & {ACC} & {F1} 
      & {ACC} & {F1} \\
      \midrule
      Zero-Shot    & 51.13 & 67.66 & 50.37 & 66.74 & 60.22 & 75.06 \\
      Reasoning    & 69.68 & 72.20 & 58.05 & 50.17 & 58.04 & 72.15 \\
      \midrule
      One-Shot     & 78.28 & 78.76 & 65.92 & 66.50 & 60.20 & 75.08 \\
      Two-Shot     & 69.23 & 74.44 & 58.29 & 42.01 & 60.22 & 75.10 \\
      Three-Shot   & 56.11 & 68.20 & 56.01 & 41.77 & 60.35 & 75.16 \\
      \midrule
      Emb+GNN      & \textbf{84.16} & \textbf{83.72} & \textbf{96.71} & \textbf{96.75} & \textbf{84.93} &  \textbf{87.99} \\
      \bottomrule
    \end{tabular}
    }
  \end{minipage}
  % \hfill
  \begin{minipage}[t]{0.49\textwidth}
    \centering
    \captionof{table}{Different LLMs on Politifact (\%).}
    \label{tab:llm}
    \scalebox{0.8}{
    \renewcommand{\arraystretch}{1.15}
    \setlength{\tabcolsep}{7pt}
    \begin{tabular}{l|cc|cc}
      \toprule
      \multirow{2}{*}{\textbf{LLM}} 
        & \multicolumn{2}{c|}{\textbf{Zero-Shot}} 
        & \multicolumn{2}{c}{\textbf{One-Shot}} \\
      & {ACC} & F1 
      & {ACC} & {F1} \\
      \midrule
      Qwen3-8B \citeyearpar{yang2025qwen3}       & 51.13 & 67.66 & 78.28 & 78.76 \\
      % Llama-3.1-8B \citeyearpar{dubey2024llama}  & 50.23 & 66.87 & 63.80 & 57.89 \\
      GPT-4.1-nano (2025)   & 51.58 & 67.87 & 57.47 & 69.68 \\
      GPT-4.1-mini (2025)   & 72.85 & 76.56 & 83.26 & 83.11 \\
      GPT-4.1      (2025)   & {84.62} & \textbf{84.40} & \textbf{85.52} & \textbf{84.16} \\
      GPT-5        (2025)   & \textbf{85.07} & 83.74 & 85.07 & 83.90 \\
      O3         (2025)   & 83.71 & 82.00 & 84.16 & 82.76 \\
      \bottomrule
    \end{tabular}
    }
  \end{minipage}
\end{table*}

\subsection{Problem Definition}

While the dataset is in the domain of fake news detection, we formally define the general task of TAG outlier detection as follows:

\begin{definition}[Text-Attributed Graph Outlier Detection]
Let $\mathbb{G} = \{\mathcal{G}_1, \mathcal{G}_2, \ldots, \mathcal{G}_N\}$ denote a collection of $N$ text-attributed graphs. Each graph $\mathcal{G}_i = (\mathcal{V}_i, \mathcal{E}_i, \mathbf{X}_i, \mathcal{T}_i)$ consists of a set of nodes $\mathcal{V}_i$, a set of edges $\mathcal{E}_i \subseteq \mathcal{V}_i \times \mathcal{V}_i$, textual node attributes $\mathbf{X}_i$, and optional node associated timestamp $\mathcal{T}_i$. Each graph is annotated with a binary label $y_i \in \{0,1\}$, indicating whether the graph is an outlier ($y_i = 1$) or not ($y_i = 0$). The objective of the task is to learn a function $f: \mathcal{G} \rightarrow \{0,1\}$ that predicts the binary label $\hat{y}_i$ for each graph $\mathcal{G}_i$.
\end{definition}

\subsection{Dataset Construction}
\label{sec:data}

To support outlier detection on TAG, we construct three fake news detection (sub-)datasets in the same format as \texttt{Dataset}\footnote{\url{https://pytorch-geometric.readthedocs.io/en/latest/generated/torch_geometric.data.Dataset.html}} in PyG, based on existing datasets. For Politifact and Gossipcop, we follow \citep{shu2020fakenewsnet} and \citep{dou2021user}, which jointly models news content and user interaction through propagation graphs. While the original datasets provided only preprocessed text embeddings via BERT or word2vec as node features, we retain the raw textual content of both the news articles and user historical posts, allowing for more flexibility in LLM-based methods. For Fakeddit, we adopt all samples in \citep{nakamura2019r} as news content, and represent comment users\footnote{\url{https://pushshift.io}} as child nodes in the graph. We filter out bot users\footnote{\url{https://botrank.pastimes.eu}} and remove the news that has no user comments. We mask out personal ID in the raw text on all subsets.

Figure \ref{fig:example} illustrates a toy example of a news propagation graph $\mathcal{G}$ in \system. 
A detailed case study of a simple real instance is provided in Appendix~\ref{appx:case}. 
The root node $v_{0}$ is the origin of the propagation graph (i.e., news itself), while the child nodes $(v_{j}, v_{k}) \in \mathcal{E}$ denote users involved in the propagation. Each node in the graph is attributed with text: the root node is attributed with the original news content $\mathbf{x}_{0}$, child nodes (users) are attributed with historical user posts $\mathbf{x}_{j}, j>0$. To constrain the text length, we limit each user to their 200 most recent posts. Ground-truth outlier labels, indicating whether the news/graph is fake (1) or real (0), are adopted from prior work. Additionally, we include the Unix timestamp for each node to capture temporal information. Preliminary experiments indicate naively put the Unix timestamp into the LLM prompt does not significantly affect detection performance. We also provide public train, validation, and test splits consistent with \citep{shu2020fakenewsnet} and \citep{nakamura2019r}.

\section{Experiments}

        In this section, we benchmark the performance of existing methods on \system. We aim to answer \textbf{RQ1} (Section \ref{sec:supervision}): How do different levels of supervision affect performance on text-attributed graph-based fake news detection? \textbf{RQ2} (Section \ref{sec:llm}): How do different LLMs vary in their detection performance? \textbf{RQ3} (Section \ref{sec:ablation}): Does each component of the text-attributed graph contribute to LLM-based fake news detection?

\subsection{Prompting vs. Embedding}
\label{sec:supervision}

We start from evaluating different levels of supervision with LLMs.

\begin{itemize}
\item \textit{Zero-shot inference}: standard \textbf{Zero-Shot} prompting and Chain-of-Thought~\textbf{Reasoning} \citep{wei2022chain};
\item \textit{Few-shot in-context learning} (ICL; \citealp{brown2020language}), prompting the LLM with a few labeled examples per class, including \textbf{One-Shot}, \textbf{Two-Shot}, and \textbf{Three-Shot};
\item \textit{Supervised learning}: training a graph neural network (GNN) on LLM-based embeddings, denoted as \textbf{Emb+GNN}, following UPFD~\citep{dou2021user}.
\end{itemize}

Details of our prompt design for graph data are provided in Appendix~\ref{sec:prompt}. 
% Details of our prompt design for graph data are provided in the code implementation. 
For a fair comparison, we adopt Qwen3-8B~\citep{yang2025qwen3} for prompting, and use Qwen3-Embedding-8B~\citep{zhang2025qwen3} as the embedding model. We implement the GNN with GraphSAGE \citep{hamilton2017inductive}. We evaluate all the performance on test set of each subset, and sample the few-shot examples from validation set. The performance comparison of accuracy (ACC) and F1 score (F1) is shown in Table~\ref{tab:main}. From the table, we have three key findings:

\textbf{1. In-context learning and reasoning help.} We observe a substantial improvement in LLM accuracy on Politifact, rising from 51.13 to 78.28 with only one-shot examples. Moreover, even without any labeled examples, LLM reasoning also improves the accuracy to 69.68. A similar trend is observed on Gossicop, though not on Fakeddit. We hypothesize that this is due to the lower overlap between the pretraining data of Qwen3-8B and Fakeddit compared to other two subsets.

\textbf{2. Supervised learning remains effective.} Emb+GNN consistently outperforms Qwen3-8B-based methods across all three subsets. The performance gap is particularly pronounced on larger datasets, highlighting the importance of abundant supervision for effective fake news detection. Furthermore, when compared to the performance of BERT embeddings with GraphSAGE reported in \citet{dou2021user}, the results using LLM-based embeddings are comparable, suggesting that the performance bottleneck lies not in the embeddings.

\textbf{3. Two-shot and three-shot learning degrade with longer context.} On Politifact and Gossicop, performance declines as the number of in-context examples increases. This degradation is not observed on Fakeddit, which has a smaller average graph size. We attribute this phenomenon to the increased context length, which may exceed the effective attention capacity of the LLM.

\subsection{Performance of Different LLMs}
\label{sec:llm}

In Table~\ref{tab:llm}, we benchmark the performance of various LLMs on Politifact in both zero-shot and one-shot settings. The results from GPT-4.1-nano to GPT-4.1 indicate that performance generally improves with increasing model size. Notably, zero-shot GPT-4.1 already surpasses the supervised Qwen3-Embedding-8B with GraphSAGE. In addition, providing just one example per class (one-shot ICL) yields a substantial boost for smaller models.

\subsection{Ablation Study}
\label{sec:ablation}

\begin{figure}[t]
    \centering
    \includegraphics[width=0.9\linewidth]{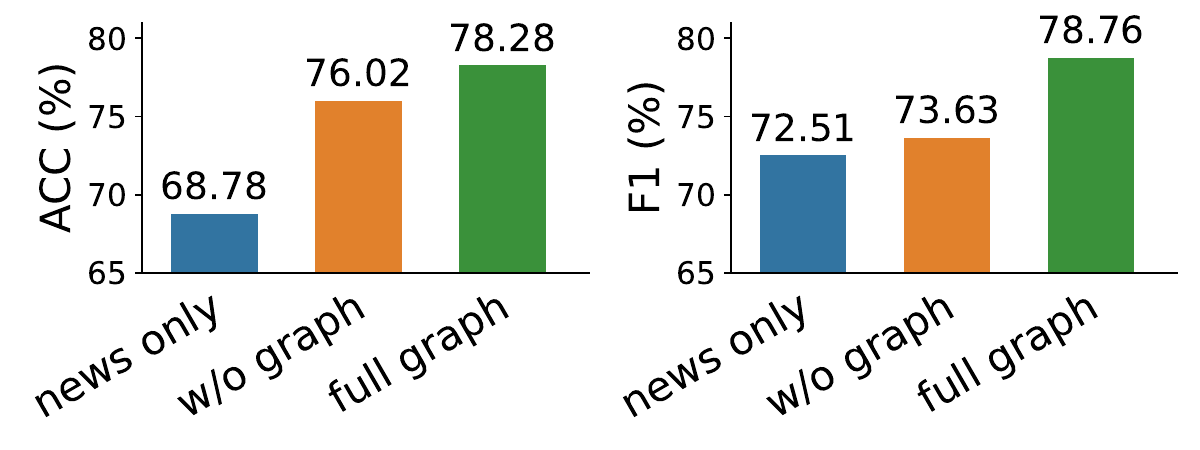}
    \caption{Ablation study of one-shot ICL on Politifact.}
    \label{fig:ablation}
\end{figure}

We further conduct an ablation study on Politifact to show the importance of text-attributed graph. We consider two variants: \textbf{w/o graph}, which removes graph structure while retaining the new content and user posts in the prompt, and \textbf{news only}, which includes only the news content in the prompt. The results, shown in Figure~\ref{fig:ablation}, indicate that the full graph yields the best performance. Removing the graph structure or user posts leads to a noticeable drop in both accuracy and F1 score, highlighting the importance of each component in TAG.

\section{Conclusion}

To address the prevailing challenge of scarce real-world datasets, we introduce \system, a text-attributed graph dataset designed for outlier detection, specifically fake news detection. \system comprises of three subsets of varying scales, enabling comprehensive benchmarking. We further evaluate a range of methods on \system to establish baseline performance. We hope this paper can provide valuable insights and facilitate future advancements in graph language models and trustworthy AI.

\section{Limitations}

As discussed in Section~\ref{sec:data}, our experiments on \system are currently limited to static graphs, without considering the timestamp on each node. While preliminary experiments suggest that naive approaches to utilizing timestamps are ineffective, we reserve the exploration of more sophisticated temporal modeling techniques for future work. Furthermore, due to resource constraints and budgetary limitations, our evaluation does not include larger open-source models with over 10B parameters, as well as other prominent LLM families such as the Claude, Gemini, Mistral, and Deepseek.

% \begin{acks}
% \end{acks}

\bibliographystyle{ACM-Reference-Format}
\bibliography{main}

@inproceedings{dou2021user,
  title={User preference-aware fake news detection},
  author={Dou, Yingtong and Shu, Kai and Xia, Congying and Yu, Philip S and Sun, Lichao},
  booktitle={Proceedings of the 44th international ACM SIGIR conference on research and development in information retrieval},
  pages={2051--2055},
  year={2021}
}

@article{yan2023comprehensive,
  title={A Comprehensive Study on Text-attributed Graphs: Benchmarking and Rethinking},
  author={Yan, Hao and Li, Chaozhuo and Long, Ruosong and Yan, Chao and Zhao, Jianan and Zhuang, Wenwen and Yin, Jun and Zhang, Peiyan and Han, Weihao and Sun, Hao and others},
  journal={Advances in Neural Information Processing Systems},
  volume={36},
  pages={17238--17264},
  year={2023}
}

@inproceedings{NEURIPS2024_a65d054a,
 author = {Zhang, Jiasheng and Chen, Jialin and Yang, Menglin and Feng, Aosong and Liang, Shuang and Shao, Jie and Ying, Rex},
 booktitle = {Advances in Neural Information Processing Systems},
 pages = {91405--91429},
 title = {DTGB: A Comprehensive Benchmark for Dynamic Text-Attributed Graphs},
 volume = {37},
 year = {2024}
}

@inproceedings{10.5555/3737916.3739865,
author = {Li, Zhuofeng and Gou, Zixing and Zhang, Xiangnan and Liu, Zhongyuan and Li, Sirui and Hu, Yuntong and Ling, Chen and Zhang, Zheng and Zhao, Liang},
title = {TEG-DB: a comprehensive dataset and benchmark of textual-edge graphs},
year = {2025},
isbn = {9798331314385},
booktitle = {Proceedings of the 38th International Conference on Neural Information Processing Systems},
articleno = {1949},
numpages = {19},
location = {Vancouver, BC, Canada},
series = {NIPS '24}
}

@article{chen2023label,
  title={Label-free node classification on graphs with large language models (llms)},
  author={Chen, Zhikai and Mao, Haitao and Wen, Hongzhi and Han, Haoyu and Jin, Wei and Zhang, Haiyang and Liu, Hui and Tang, Jiliang},
  journal={arXiv preprint arXiv:2310.04668},
  year={2023}
}

@article{yasunaga2022linkbert,
  title={Linkbert: Pretraining language models with document links},
  author={Yasunaga, Michihiro and Leskovec, Jure and Liang, Percy},
  journal={arXiv preprint arXiv:2203.15827},
  year={2022}
}

@article{xu2025graph,
  title={Graph synthetic out-of-distribution exposure with large language models},
  author={Xu, Haoyan and Yao, Zhengtao and Wang, Ziyi and Cheng, Zhan and Hu, Xiyang and Li, Mengyuan and Zhao, Yue},
  journal={arXiv preprint arXiv:2504.21198},
  year={2025}
}

@inproceedings{zhu2025graphclip,
  title={Graphclip: Enhancing transferability in graph foundation models for text-attributed graphs},
  author={Zhu, Yun and Shi, Haizhou and Wang, Xiaotang and Liu, Yongchao and Wang, Yaoke and Peng, Boci and Hong, Chuntao and Tang, Siliang},
  booktitle={Proceedings of the ACM on Web Conference 2025},
  pages={2183--2197},
  year={2025}
}

@article{li2023survey,
  title={A survey of graph meets large language model: Progress and future directions},
  author={Li, Yuhan and Li, Zhixun and Wang, Peisong and Li, Jia and Sun, Xiangguo and Cheng, Hong and Yu, Jeffrey Xu},
  journal={arXiv preprint arXiv:2311.12399},
  year={2023}
}

@article{chen2024exploring,
  title={Exploring the potential of large language models (llms) in learning on graphs},
  author={Chen, Zhikai and Mao, Haitao and Li, Hang and Jin, Wei and Wen, Hongzhi and Wei, Xiaochi and Wang, Shuaiqiang and Yin, Dawei and Fan, Wenqi and Liu, Hui and others},
  journal={ACM SIGKDD Explorations Newsletter},
  volume={25},
  number={2},
  pages={42--61},
  year={2024},
  publisher={ACM New York, NY, USA}
}

@article{yang2021graphformers,
  title={Graphformers: Gnn-nested transformers for representation learning on textual graph},
  author={Yang, Junhan and Liu, Zheng and Xiao, Shitao and Li, Chaozhuo and Lian, Defu and Agrawal, Sanjay and Singh, Amit and Sun, Guangzhong and Xie, Xing},
  journal={Advances in Neural Information Processing Systems},
  volume={34},
  pages={28798--28810},
  year={2021}
}

@inproceedings{xiao2020timme,
  title={Timme: Twitter ideology-detection via multi-task multi-relational embedding},
  author={Xiao, Zhiping and Song, Weiping and Xu, Haoyan and Ren, Zhicheng and Sun, Yizhou},
  booktitle={Proceedings of the 26th ACM SIGKDD international conference on knowledge discovery \& data mining},
  pages={2258--2268},
  year={2020}
}

@article{liu2023data,
  title={Data augmentation for supervised graph outlier detection with latent diffusion models},
  author={Liu, Kay and Zhang, Hengrui and Hu, Ziqing and Wang, Fangxin and Yu, Philip S},
  journal={arXiv preprint arXiv:2312.17679},
  year={2023}
}

@article{liu2022bond,
  title={Bond: Benchmarking unsupervised outlier node detection on static attributed graphs},
  author={Liu, Kay and Dou, Yingtong and Zhao, Yue and Ding, Xueying and Hu, Xiyang and Zhang, Ruitong and Ding, Kaize and Chen, Canyu and Peng, Hao and Shu, Kai and others},
  journal={Advances in Neural Information Processing Systems},
  volume={35},
  pages={27021--27035},
  year={2022}
}

@article{li2025taco,
  title={TACO: Enhancing Multimodal In-context Learning via Task Mapping-Guided Sequence Configuration},
  author={Li, Yanshu and Yun, Tian and Yang, Jianjiang and Feng, Pinyuan and Huang, Jinfa and Tang, Ruixiang},
  journal={arXiv preprint arXiv:2505.17098},
  year={2025}
}

@article{yang2025qwen3,
  title={Qwen3 technical report},
  author={Yang, An and Li, Anfeng and Yang, Baosong and Zhang, Beichen and Hui, Binyuan and Zheng, Bo and Yu, Bowen and Gao, Chang and Huang, Chengen and Lv, Chenxu and others},
  journal={arXiv preprint arXiv:2505.09388},
  year={2025}
}

@article{hamilton2017inductive,
  title={Inductive representation learning on large graphs},
  author={Hamilton, Will and Ying, Zhitao and Leskovec, Jure},
  journal={Advances in neural information processing systems},
  volume={30},
  year={2017}
}

@inproceedings{fatemitalk,
  title={Talk like a Graph: Encoding Graphs for Large Language Models},
  author={Fatemi, Bahare and Halcrow, Jonathan and Perozzi, Bryan},
  booktitle={The Twelfth International Conference on Learning Representations},
  year={2024}
}

@article{shu2020fakenewsnet,
  title={Fakenewsnet: A data repository with news content, social context, and spatiotemporal information for studying fake news on social media},
  author={Shu, Kai and Mahudeswaran, Deepak and Wang, Suhang and Lee, Dongwon and Liu, Huan},
  journal={Big data},
  volume={8},
  number={3},
  pages={171--188},
  year={2020},
  publisher={Mary Ann Liebert, Inc., publishers 140 Huguenot Street, 3rd Floor New~…}
}

@article{nakamura2019r,
  title={r/fakeddit: A new multimodal benchmark dataset for fine-grained fake news detection},
  author={Nakamura, Kai and Levy, Sharon and Wang, William Yang},
  journal={arXiv preprint arXiv:1911.03854},
  year={2019}
}

@article{wang2017liar,
  title={" liar, liar pants on fire": A new benchmark dataset for fake news detection},
  author={Wang, William Yang},
  journal={arXiv preprint arXiv:1705.00648},
  year={2017}
}

@article{yang2024ad,
  title={Ad-llm: Benchmarking large language models for anomaly detection},
  author={Yang, Tiankai and Nian, Yi and Li, Shawn and Xu, Ruiyao and Li, Yuangang and Li, Jiaqi and Xiao, Zhuo and Hu, Xiyang and Rossi, Ryan and Ding, Kaize and others},
  journal={arXiv preprint arXiv:2412.11142},
  year={2024}
}

@article{brown2020language,
  title={Language models are few-shot learners},
  author={Brown, Tom and Mann, Benjamin and Ryder, Nick and Subbiah, Melanie and Kaplan, Jared D and Dhariwal, Prafulla and Neelakantan, Arvind and Shyam, Pranav and Sastry, Girish and Askell, Amanda and others},
  journal={Advances in neural information processing systems},
  volume={33},
  pages={1877--1901},
  year={2020}
}

@article{wei2022chain,
  title={Chain-of-thought prompting elicits reasoning in large language models},
  author={Wei, Jason and Wang, Xuezhi and Schuurmans, Dale and Bosma, Maarten and Xia, Fei and Chi, Ed and Le, Quoc V and Zhou, Denny and others},
  journal={Advances in neural information processing systems},
  volume={35},
  pages={24824--24837},
  year={2022}
}

@article{zhang2025qwen3,
  title={Qwen3 Embedding: Advancing Text Embedding and Reranking Through Foundation Models},
  author={Zhang, Yanzhao and Li, Mingxin and Long, Dingkun and Zhang, Xin and Lin, Huan and Yang, Baosong and Xie, Pengjun and Yang, An and Liu, Dayiheng and Lin, Junyang and others},
  journal={arXiv preprint arXiv:2506.05176},
  year={2025}
}

@article{li2024nlp,
  title={Nlp-adbench: Nlp anomaly detection benchmark},
  author={Li, Yuangang and Li, Jiaqi and Xiao, Zhuo and Yang, Tiankai and Nian, Yi and Hu, Xiyang and Zhao, Yue},
  journal={arXiv preprint arXiv:2412.04784},
  year={2024}
}

\appendix

\section{Data Instance}
\label{appx:case}

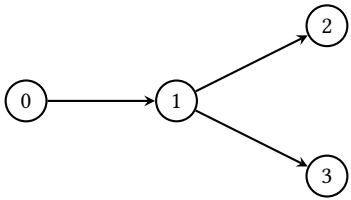
\begin{figure}[ht]
\centering
\begin{tikzpicture}[->,>=stealth,thick,node distance=2cm]
  \node[circle,draw] (A) at (0,0) {0};
  \node[circle,draw] (B) at (2,0) {1};
  \node[circle,draw] (C) at (4,1) {2};
  \node[circle,draw] (D) at (4,-1) {3};

  \draw (A) -- (B);
  \draw (B) -- (C);
  \draw (B) -- (D);
\end{tikzpicture}
\caption{The graph structure of the data instance.}
\label{fig:case}
\end{figure}

To provide a case study on \system, we illustrate an instance of news propagation graph from Politifact in Figure~\ref{fig:case}. This graph has 4 nodes (from Node 0 to Node 3) and 3 edges. The corresponding raw textual attributes for each node are as follows:

\begin{lstlisting}[basicstyle=\ttfamily\small, breaklines=true, frame=single]
Node 0 (News Content): Based on the Monthly Treasury Statement for August and the Daily Treasury Statements for September. CBO estimates that the federal budget deficit was about $1.30 trillion in fiscal year 2011, approximately the same dollar amount as the shortfall recorded in 2010. The 2011 deficit was equal to 8.6 percent of gross domestic product, CBO estimates, down from 8.9 percent in 2010 and 10.0 percent in 2009, but greater than in any other year since 1945. The estimated 2011 total reflects the shift of some payments from fiscal year 2012 into fiscal year 2011 (that is, from October to September, because October 1 fell on a weekend); without that shift, the deficit in 2011 would have been $1.27 trillion. CBO's deficit estimate is based on data from the Daily Treasury Statements; the Treasury Department will report the actual deficit for fiscal year 2011 later this month.

Node 1 (User Post): Teri and I wish you a Merry Christmas!  Wishing peace and joy to my friends and neighbors in the Jewish community on this first night of Hanukkah ...

Node 2 (User Post): RT @user: LATE BREAKING: This morning the FBI arrested a member of the Cincinnati city council for accepting bribe money in exch ...

Node 3 (User Post): .@user There they go again, with superficial over #Substance. One wd suggest they Try to compare their er ...
\end{lstlisting}

\section{Prompt Design}
\label{sec:prompt}

To enable LLM-based fake news detection on text-attributed graphs, we design a structured prompt to encode both the news content and the associated propagation graph for LLMs, following graph prompt in \citep{fatemitalk}. The prompt includes a system prompt and a user prompt.

\subsection{System Prompt}
The system prompt sets the context for the LLM, describing the task and the format of the input. The following system prompt is used for experiments:

\begin{tcolorbox}[
    colback=gray!4,     % Background color
    colframe=black!55,% Frame color
    title=System Prompt
]
You are a fake news detection assistant analyzing news propagation graph on social networks.\\

You will be provided with:\\
- the content of a news (corresponding to the root node in the propagation graph),\\
- user posts (each corresponding to a subsequent node in the propagation graph),\\
- the structure of the propagation graph. The edges indicate the propagation relationships.\\

Based on the content and graph structure, your task is to determine whether the news is `Real` or `Fake`.\\

Output: respond with only the fake news classification label: `Real` or `Fake`.
\end{tcolorbox}

\subsection{User Prompt}

The user prompt encodes the instance to be classified, including the news content, user posts, and the graph structure. For few-shot in-context learning, a few labeled examples are included in the same format, each followed by the correct output label (`Real` or `Fake`). In experiments, to fit the prompt into the context window, we restrict the post content of each user to 500 characters and the maximum number of users to 30. Below is a demostration of the prompt provided to the LLM:

\begin{tcolorbox}[
    colback=gray!4,     % Background color
    colframe=black!55,% Frame color
    title=User Prompt
]
EXAMPLES:\\

Input:\\
Node 0 (NEWS): <news content>\\
Node 1 (USER POST): <user post>\\
Node 2 (USER POST): <user post>\\

Graph Structure:\\
Node 0 propagate to Node 1,\\
Node 0 propagate to Node 2\\

Output: Real\\

Input:\\
Node 0 (NEWS): <news content>\\
Node 1 (USER POST): <user post>\\
Node 2 (USER POST): <user post>\\

Graph Structure:\\
Node 0 propagate to Node 1,\\
Node 1 propagate to Node 2\\

Output: Fake\

END OF EXAMPLES. Classify the following news:\\

Input:\\
Node 0 (NEWS): <news content>\\
Node 1 (USER POST): <user post>\\
Node 2 (USER POST): <user post>\\

Graph Structure:\\
Node 0 propagate to Node 1,\\
Node 1 propagate to Node 2\\
\end{tcolorbox}

\end{document}